\documentclass[showpacs,preprintnumbers,amsmath,amssymb,preprint]{revtex4}
\usepackage{amsmath,amssymb}
\usepackage{epsfig}
\usepackage{citesort}
\usepackage{graphicx}
\begin{document}
%-------------------------------------------------------------------------------------
%TITLE
%-------------------------------------------------------------------------------------
\title{Controlled transport of solitons and bubbles using external perturbations}
%-------------------------------------------------------------------------------------
%AUTHORS
%-------------------------------------------------------------------------------------
\author{J. A. Gonz\'alez$^{1}$\footnote{E-mail: jorge@ivic.ve}}
%-------------------------------------------------------------------------------------
\author{A. Marcano$^{1}$}%\footnote{E--mail: marcano@ivic.ve}}
%-------------------------------------------------------------------------------------
\author{B. A. Mello$^{2}$}%\footnote{E--mail: bmello@us.ibm.com}}
%-------------------------------------------------------------------------------------
\author{L. Trujillo$^{1}$\footnote{E--mail: leo@ivic.ve}}
%-------------------------------------------------------------------------------------
%AFFILIATIONS
%-------------------------------------------------------------------------------------
\affiliation{$^1$Centro de F\'{\i}sica, Instituto Venezolano de
Investigaciones Cient\'{\i}ficas (IVIC), A.P. 21827, Caracas
1020--A, Venezuela}
%-------------------------------------------------------------------------------------
%\affiliation{$^2$PMMH,
%Physique et M\'ecanique des Milieux H\'et\'erog\`{e}nes,
%UMR CNRS 7636,
%ESPCI, 10 rue Vauquelin, 75231 Paris Cedex 05, France}
%-------------------------------------------------------------------------------------
\affiliation{$^2$IBM Corporation, Thomas I. Watson Research
Center, Yorktown Heights, NY 10598, USA}
%-------------------------------------------------------------------------------------
\date{\today}
%-------------------------------------------------------------------------------------
%ABSTRACT
%-------------------------------------------------------------------------------------
\begin{abstract}
We investigate generalized soliton--bearing systems in the
presence of external perturbations. We show the possibility of the
transport of solitons using external waves, provided the waveform
and its velocity satisfy certain conditions. We also investigate
the stabilization and transport of bubbles using external
perturbations in $3D$--systems. We also present the results of
real experiments with laser--induced vapor bubbles in liquids.
\end{abstract}
%-------------------------------------------------------------------------------------
%-------------------------------------------------------------------------------------
%\preprint{Chaos, Solitons and Fractal}
%-------------------------------------------------------------------------------------
\maketitle
%-------------------------------------------------------------------------------------

%-------------------------------------------------------------------------------------
%MAIN TEXT
%-------------------------------------------------------------------------------------

%-------------------------------------------------------------------------------------
\section{Introduction}
%-------------------------------------------------------------------------------------
Recently, in a very interesting paper, Zheng {\it et al.,}
\cite{1:ZCH02} have studied the collective directed transport of
symmetrically coupled lattices in  symmetric periodic potentials.
They show that under the action of an external wave, that breaks
the spatiotemporal symmetry and introduces inhomogeneities of the
lattice, a net unidirectional current can be observed. Apparently
the current originates from the collaboration of the lattice and
the wave (amplitude, frequency, and phase shifts). The study of
directed transport of particles is very important in the physics
of molecular motors \cite{2:As97,3:Ju97}, vortex dynamics in
superconductors \cite{4:Lee99,5:Ols01}, Josephson junction
lattices \cite{6:Tri00}, nanotechnology \cite{7:Rou94,8:Oud99} and
many other systems.

Many studies have been dedicated to directed transport in
spatiotemporal systems
\cite{9:Rei02,10:DV95,11:Csa97,12:Iga01,13:Zhe01,14:ZL01,15:Jun96,16:Mat00,17:Fla00,18:Yev01}.
These include systems with deterministic ac drives and stochastic
forces
\cite{13:Zhe01,14:ZL01,15:Jun96,16:Mat00,17:Fla00,18:Yev01}.

Very important are models composed of a symmetrically coupled
lattice in a symmetric potential field, which is driven by an
external wave \cite{1:ZCH02}.

The equation studied by Zheng {\it et al.,} \cite{1:ZCH02} is the
following:
%-------------------------------------------------------------------------------------
\begin{equation}
\dot{\theta}_j = - d \sin \theta_j + \kappa \left( \theta_{j+1} -
2 \theta_j + \theta_{j-1} \right) + A \cos \left( \omega t + j\phi
\right).
\label{Eq1:FK}
\end{equation}
%-------------------------------------------------------------------------------------

This is a Frenkel--Kontorova model  driven by the wave
$A\cos\left( \omega t + j\phi \right)$.

The average current introduced by them $J = \lim_{T\rightarrow NT}
\sum_{j=1}^{N}\int_0^T \dot{\theta}_i (t) \mathrm{d}t$ was shown
to be nonzero for certain values of the parameters of the system.

We wish to re-state this problem in terms of the directed motion
of solitons. Solitons are considered as particles in many systems.
Moreover, in many systems, the solitons are charge carriers
\cite{19:GK92,20:HKSS88,21:Dav84,22:Ric79,23:Lu88}. So it is very
important to study the directed transport of these objects.

In the present paper we investigate the generalized Klein--Gordon
equation:
%-------------------------------------------------------------------------------------
\begin{equation}
\phi_{tt} + \gamma \phi_t - \phi_{xx} - G(\phi) = F(x-wt),
\label{Eq2:KG}
\end{equation}
%-------------------------------------------------------------------------------------
where $G(\phi)= - \partial U(\phi)/\partial\phi$, $U(\phi)$ is a
symmetric potential with, at least, two minima at $\phi = \phi_1$,
$\phi = \phi_3$, and a maximum at $\phi = \phi_2$  ($\phi_1 <
\phi_2 < \phi_3$); $F(x - wt)$ describes an external wave moving
with velocity $w$. Both the sine--Gordon and $\phi^4$ equations
are particular cases of equation (\ref{Eq2:KG}). The sine--Gordon
equation is a continuous relative of Eq.~(\ref{Eq1:FK}).

These equations possess important applications in condensed matter
physics. For instance, in solid sate physics, they describe domain
walls in ferromagnets and ferroelectric materials, dislocations in
crystals, charge--density waves, interphase boundaries in metal
alloys, fluxons in long Josephson junctions and Josephson
transmission lines, etc \cite{24:LS78,25:BkT80,26:KM89}.

Our initial condition will be a kink--soliton whose
center--of--mass is situated at rest in the point $x=0$. We will
show that the transport of the soliton by the wave depends on the
shape of the wave and the parameters of the system.

We will also investigate the stabilization and transport of
bubbles using external perturbations in $3D$--systems as the
following
%-------------------------------------------------------------------------------------
\begin{equation}
\phi_t - \nabla^2\phi - G(\phi) = F (x,y,z,t).
\label{Eq3:Sys3d}
\end{equation}
%-------------------------------------------------------------------------------------

We will also present the results of real experiments with
laser--induced bubbles in liquids.

The bubbles can be trapped, stabilized and transported by a laser
beam of relatively low power. We will show that the directed
transport of solitons and bubbles are related. For both phenomena
some conditions for the external perturbations must be satisfied.

%-------------------------------------------------------------------------------------
\section{Stability and dynamics of solitons}
%-------------------------------------------------------------------------------------
Before discussing equation (\ref{Eq2:KG}) we would like to present
some results about the dynamics of solitons in the presence of
time--independent inhomogeneous  perturbations
\cite{27:GMRG98,28:GH92,29:GM96,30:GGB96,31:GBG99,32:GBG02} as in
the following equation:
%-------------------------------------------------------------------------------------
\begin{equation}
\phi_{tt} + \gamma \phi_t - \phi_{xx} - G(\phi) = F(x).
\label{Eq4:tiifs}
\end{equation}
%-------------------------------------------------------------------------------------

Suppose we are interested in the stability of  a soliton situated
in equilibrium positions created by the inhomogeneous force
$F(x)$. We construct a solution $\phi_k(x)$ with the topological
and general properties of a kink. Then, we solve an inverse
problem such  that $F(x)$ possesses the properties of the physical
system we are studying
\cite{27:GMRG98,28:GH92,29:GM96,30:GGB96,31:GBG99,32:GBG02}. Now
we can investigate the stability of the solution.

For this we should solve the spectral problem
%-------------------------------------------------------------------------------------
\begin{equation}
\hat{\mathrm{L}} f(x) = \Gamma f(x),
\label{Eq5:SP}
\end{equation}
%-------------------------------------------------------------------------------------
where $\hat{\mathrm{L}} = -\partial_{xx} - \left.\frac{\partial
G(\phi)}{\partial \phi}\right|_{\phi=\phi_k(x)}$, $\phi(x,t)=
\phi_k(x) + f(x) \mathrm{e}^{\lambda t}$, $\Gamma = -(\lambda^2 +
\gamma \lambda)$.

%-------------------------------------------------------------------------------------

The results obtained with this function can be generalized
qualitatively to other systems topologically equivalent to the
exactly solvable systems \cite{27:GMRG98,33:GO99}.

Let us see the following example:
%-------------------------------------------------------------------------------------
\begin{equation}
\phi_{tt}+ \gamma \phi_t - \phi_{xx} + \sin\phi = F_1(x),
\label{Eq6:Ex}
\end{equation}
%-------------------------------------------------------------------------------------
where $F_1(x) = 2(B^2 - 1)\sinh(Bx)/\cosh^2(Bx)$.

The exact stationary solution of Eq.~(\ref{Eq6:Ex}) is
$\phi_k(x)=4\arctan[\exp(Bx)]$. This solution represents a
kink--soliton equilibrated on the position $x=0$. The stability
problem (\ref{Eq5:SP}) can be solved exactly. The eigenvalues of
the discrete spectrum are given by the formula
%-------------------------------------------------------------------------------------
\begin{equation}
\Gamma_n = B^2(\Lambda + 2\Lambda n -n^2) -1,
\label{Eq7:DS}
\end{equation}
%-------------------------------------------------------------------------------------
where $\Lambda (\Lambda+1) = 2/B^2$. The integer part of
$\Lambda$, i.e., $[\Lambda]$, yields the number of eigenvalues in
the discrete spectrum, which correspond to the soliton modes (this
includes the translational mode $\Gamma_0$, and the internal or
shape modes $\Gamma_n$ ($n>0$)).

Everything that happens with this solution near point $x = 0$ can
be obtained from Eq.~(\ref{Eq7:DS}).

Now let us return to the general case (\ref{Eq4:tiifs}) and a
review of the obtained results about the dynamics of solitons in
Eq.~(\ref{Eq4:tiifs}).

Wherever $F(x)$ is positive, the kink--soliton will be accelerated
to the ``left''. Wherever $F(x)$ is negative, the kink--soliton
will be accelerated to the ``right''. The zeroes of $F(x)$ are
candidates for equilibrium positions for the kink--soliton
\cite{27:GMRG98,28:GH92,29:GM96,30:GGB96,31:GBG99,32:GBG02}. If
$F(x)$ possesses only one zero $x_0^*$ ($F(x_0^*)=0$), then it is
a stable position for the soliton if $\left. \frac{\partial
F}{\partial x}\right|_{x_0^*}>0$. Otherwise, the position is an
unstable equilibrium. The opposite is true for the
antikink--soliton. A soliton can be trapped by a $F(x)$ with a
zero that represents a stable position, and it can move away from
an unstable position.

Now we will discuss what happens in the Eq.~(\ref{Eq2:KG}) with
different waveforms.

%--------------------------------------------------------------------------------
\section{Transport of solitons using external perturbations}
%--------------------------------------------------------------------------------

%-------------------------------------------------------------------------------------
\begin{figure}
\centering
\includegraphics[width=14.0cm]{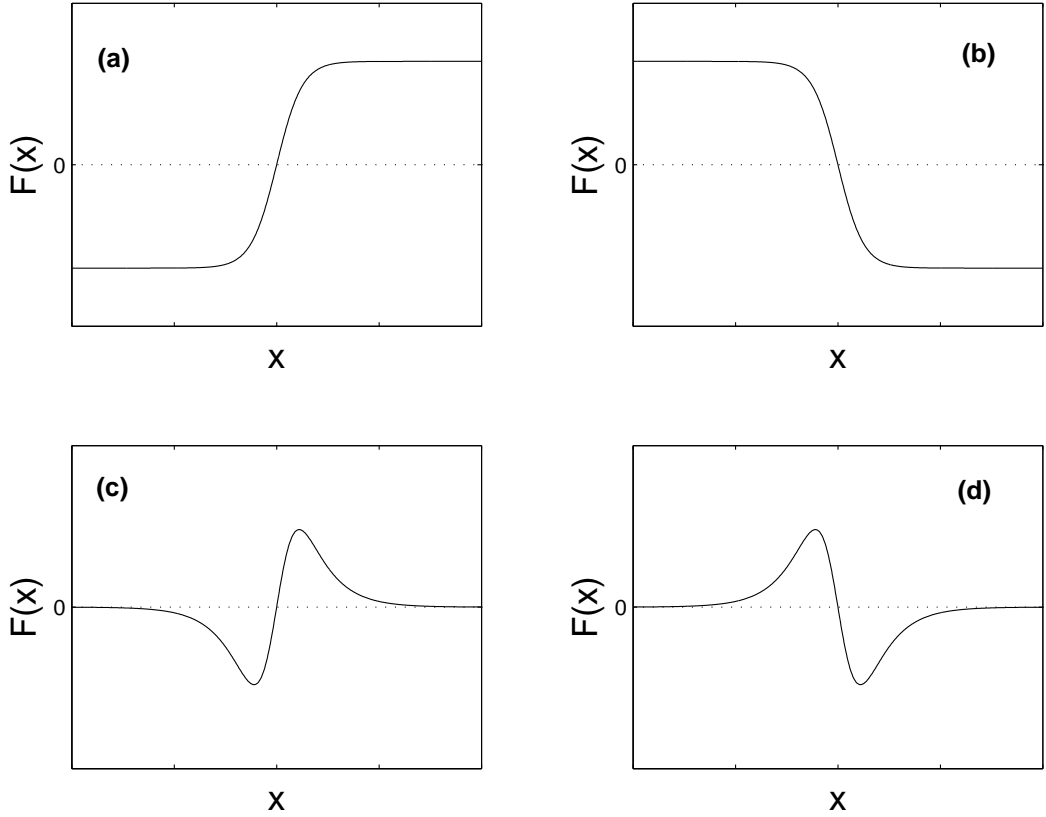}
\caption{Different waveforms for the transport of solitons in
Eq.~(\ref{Eq2:KG}). a) The soliton feels a moving deep stable
potential well. b) The soliton feels an unstable equilibrium
position. c) The soliton feels a potential well with finite walls.
d) The soliton feels a potential  with a finite maximum.}
\label{fig:Fig1}
\end{figure}
%-------------------------------------------------------------------------------------

Fig.~\ref{fig:Fig1} shows different waveforms for $F(x-vt)$. In
case a) the kink will move trapped inside the potential well
created by $F(x-vt)$. The effective potential $V(x)$ of this force
$F(x)$ possesses the property that $V(x)\rightarrow \infty$ when
$x\rightarrow \pm\infty$. Thus, the kink will move inside a
potential equivalent to a harmonic oscillator. So a wave like this
will always transport the soliton to the ``right''.

In the case b), the wave will ``push'' the soliton to the
``right'' as long as the center--of--mass of the soliton is
situated righter than the zero of $F(x)$. In fact, $V(x)$ has an
absolute maximum. However, there is a limit for the wave velocity.
If the wave velocity exceeds this limit, the center--of--mass of
the soliton can reach the maximum of the potential $V(x)$, and the
center--of--mass of the soliton will be behind the maximum of the
potential. In this situation the wave can no longer push the
soliton to the ``right''.

We will here introduce the concept of ``controlled transport''
when the soliton can be transported by the external wave and the
soliton velocity can be approximately equal to the wave velocity.
In the case a) of Fig.~\ref{fig:Fig1}, we have a controlled
transport. In the case b), the soliton is pushed by the wave but
the soliton will move ahead of the wave in an ``uncontrolled''
fashion.

\begin{figure}
\centering
\includegraphics[width=14.0cm]{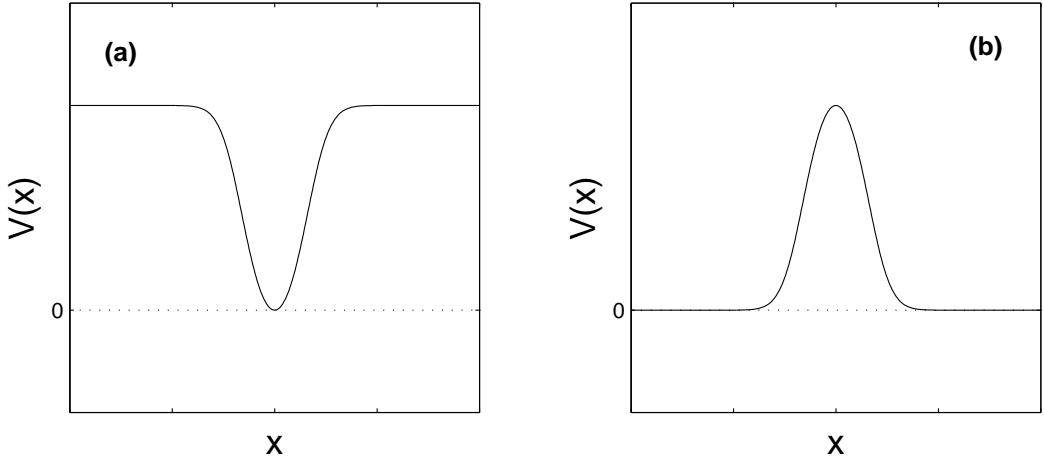}
\caption{Effective potential ``created'' by the forces shown in
Fig.~\ref{fig:Fig1} (c) and (d) respectively.}
\label{fig:Fig2}
\end{figure}

The waveform shown in Fig.~\ref{fig:Fig1}~c represents a force
that creates a stable equilibrium for the soliton. So, as in case
a), this wave can carry the soliton. However, the effective
potential corresponding to this force is finite for
$x\rightarrow\pm\infty$ (See Fig.~\ref{fig:Fig2}~a).

%-------------------------------------------------------------------------------------
%-------------------------------------------------------------------------------------

So here there is also a limit for the wave velocity.

The wave represented in Fig.~\ref{fig:Fig1}~d is an unstable
equilibrium for the soliton as in case b). This wave can push the
soliton to the ``right'' provided its velocity is smaller that
certain finite value.

%-------------------------------------------------------------------------------------
\begin{figure}
\centering
\includegraphics[width=14.0cm]{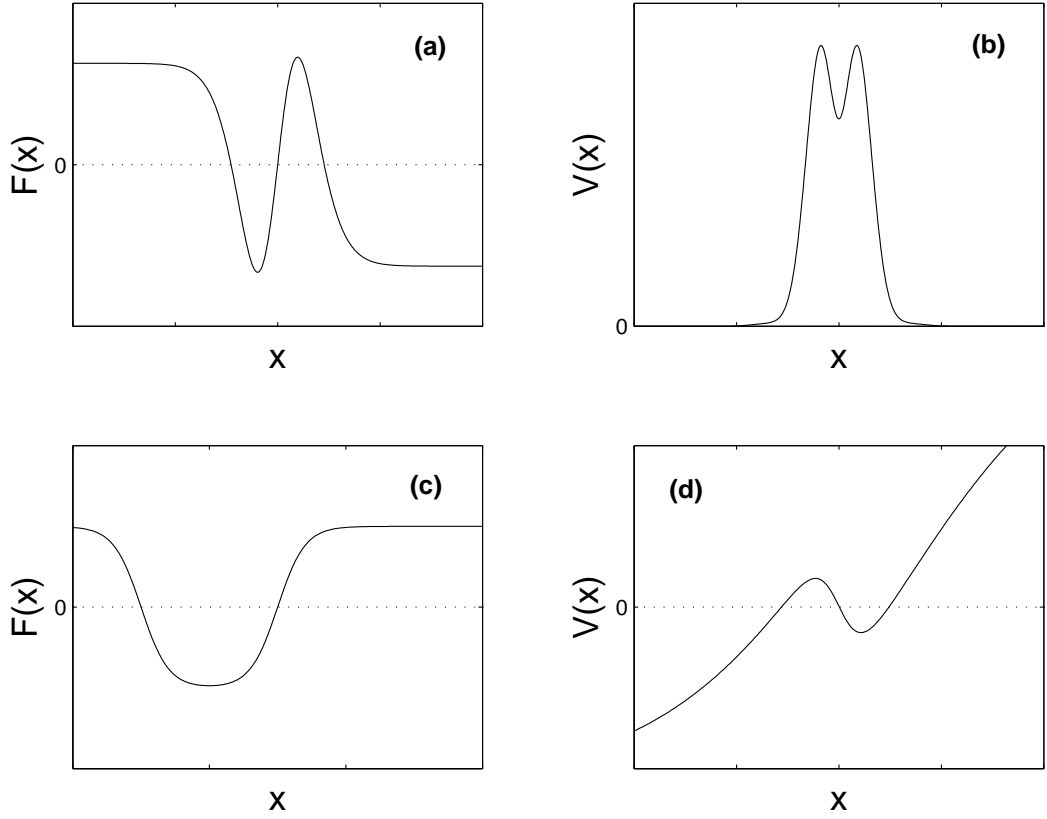}
\caption{Perturbation with several zeroes.}
\label{fig:Fig3}
\end{figure}
%-------------------------------------------------------------------------------------

Fig.~\ref{fig:Fig3} shows some different localized wave forms.

We would think that these waves could also be used to transport a
soliton in a controlled fashion because there is a stable
equilibrium involved (provided the wave velocity is smaller than
certain limit). However, here there are some additional
limitations in the shape of the wave.

Investigating different functions in (\ref{Eq4:tiifs}) and the
stability problem (\ref{Eq5:SP}), we have found that if $F(x)$
possesses several zeroes and the distance between them is smaller
than the soliton width, surprising phenomena can occur. For
instance, in the presence of forces as that shown in
Fig.~\ref{fig:Fig3}~a, if the zeroes of $F(x)$ are too close, the
center zero is not a stable point anymore. In practice, the
soliton does not ``feel'' a minimum there. It feels an unstable
equilibrium point. In this case, this wave would not be able to
transport the soliton in a controlled fashion, but it can push the
soliton to the right under certain conditions.

On the other hand, the force shown in Fig.~\ref{fig:Fig3}~c can
carry the soliton if the zeroes are sufficiently separated.
Otherwise, the soliton would not ``feel'' the left maximum of the
potential $V(x)$ (Fig.~\ref{fig:Fig3}~d). So, using this wave the
soliton cannot move trapped in the minimum.

\begin{figure}
\centering
\includegraphics[width=14.0cm]{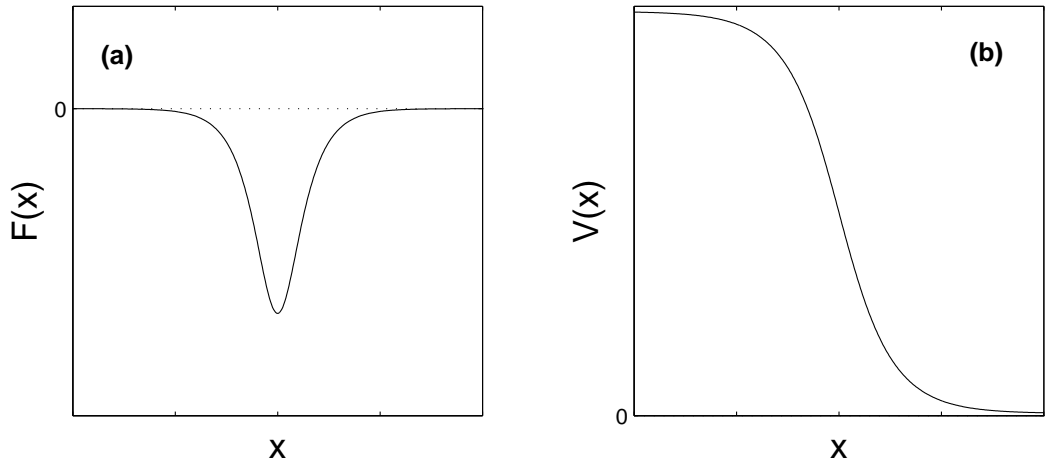}
\caption{Perturbation equivalent to a wall.}
\label{fig:Fig4}
\end{figure}

Even localized forces without a zero like that represented in
Fig.~\ref{fig:Fig4} can ``push'' the soliton to the ``right''.

%-------------------------------------------------------------------------------------
%-------------------------------------------------------------------------------------

Note that, considering the effective potential $V(x)$, this wave
``acts'' on the soliton as a moving ``wall''. But once and again,
there is a limit for the velocity of the wave.

Later we will present several examples of systems of type
(\ref{Eq2:KG}) for which we can calculate all these limit
velocities.

%-------------------------------------------------------------------------------------
\begin{figure}
\centering
\includegraphics[width=7.0cm]{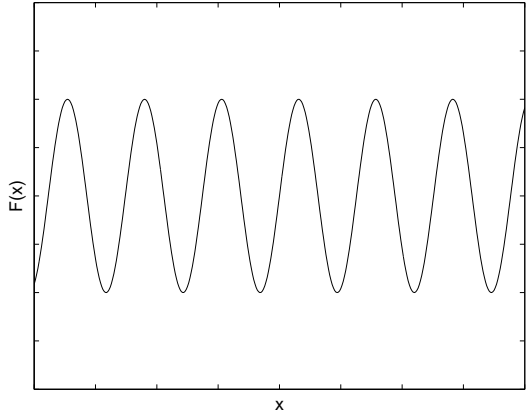}
\caption{Sinusoidal wave.}
\label{fig:Fig5}
\end{figure}
%-------------------------------------------------------------------------------------

Before, we should discuss the most ubiquitous kind of waves: the
sinusoidal wave (See Fig.~\ref{fig:Fig5}):
%-------------------------------------------------------------------------------------
\begin{equation}
F(x-wt)= A\sin[k(x-wt)].
\label{Eq8:SW}
\end{equation}
%-------------------------------------------------------------------------------------

The previous discussion of waves with the shape shown in
Fig.~\ref{fig:Fig3} leads us to think that the parameters $A$, $k$
and $w$ should satisfy some conditions for a soliton to be
transported by wave (\ref{Eq8:SW}). In fact, if the velocity $w$
is larger than some critical value, the soliton will be left
behind. If $k$ is too large (and $A$ is small), the distance
between the zeroes of $F(x)$ will be very small and the potential
wells, where the soliton can be trapped, would be so shallow that,
already with certain small velocity $w$, the soliton will be left
behind. Thus there are waves of type (\ref{Eq8:SW}) that cannot
carry the soliton.

Let us present here some concrete examples.

\begin{figure}
\centering
\includegraphics[width=7.0cm,height=8cm]{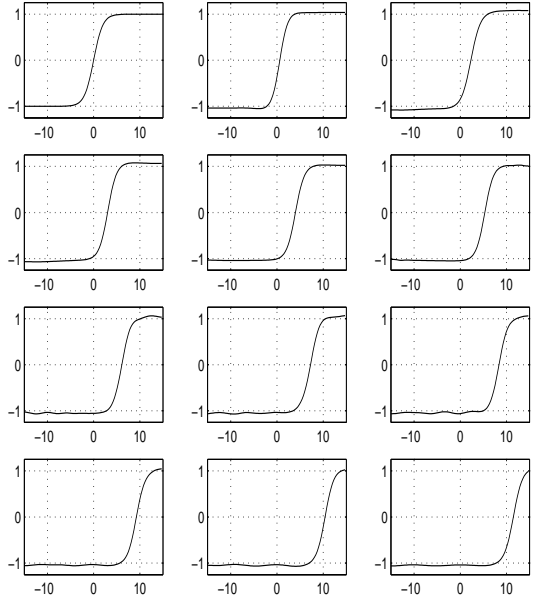}
\caption{The soliton is transported by the moving perturbation
$F(x - wt)= A\tanh[B(x-wt)]$ (See Eq.~(\ref{Eq9:phi4})).}
\label{fig:Fig6}
\end{figure}

Consider the following perturbed $\phi^4$ equation:
%-------------------------------------------------------------------------------------
\begin{equation}
\phi_{tt} + \gamma\phi_t - \phi_{xx}-\frac{1}{2}\phi +
\frac{1}{2}\phi^3 = F(x-wt),
\label{Eq9:phi4}
\end{equation}
%-------------------------------------------------------------------------------------
where $F(x-wt)= A \tanh[B(x-wt)]$, $A>0$, $B>0$. As can be
expected from our discussion, this external wave is able to
transport a soliton to the right in a controlled fashion. This can
be observed in Fig.~\ref{fig:Fig6}.
%-------------------------------------------------------------------------------------
%-------------------------------------------------------------------------------------

Later we will see that $F(x-wt)$ can be in practice not only a
wave but a controlling external perturbation. If this external
perturbation can be manipulated at wish in such a way that the
velocity $w(t)$ can be any function of time that we desire, then
we can move the soliton and place it wherever we wish.

Suppose now that in Eq.~(\ref{Eq9:phi4}) $F(x-wt)=
-A\tanh[B(x-wt)]$, $A>0$, $B>0$. In this case, the soliton is
pushed to the ``right'' while it is in the ``zone'' where
$F(x-wt)<0$ (i.e., while the center--of--mass of the soliton is
situated righter than the zero of $F(x-wt)$.) Let us define
$\phi_1<\phi_2<\phi_3$ as the  roots of the algebraic equation
$\phi-\phi^3= 2A$. The maximum velocity $v_m$ that the solution
can reach in its motion to the right \cite{28:GH92} can be
calculated using the equation
%-------------------------------------------------------------------------------------
\begin{equation}
\frac{\gamma v_m}{\sqrt{1-v_m^2}} = \frac{\phi_2}{2}
\label{Eq10:VM}
\end{equation}
%-------------------------------------------------------------------------------------

If the external wave velocity is such that $w>v_m$, then the
soliton eventually will start to move backward.

Now we will analyze the following wave $F(y)$, where $y=x-wt$:
%-------------------------------------------------------------------------------------
\begin{eqnarray}
 F(y)  =   F_1(y), & \mbox{  for $y>y*$}
\label{Eq11}\\
F(y)  =  c, & \mbox{  for $y<y*$}
\label{Eq12}
\end{eqnarray}
%-------------------------------------------------------------------------------------
where $F_1(y) = \frac{1}{2}A\tanh(By)[(A^2 - 1) + (4B^2 -
A^2)\mathrm{sech}^2(By)]$, $y*$ is the point where $F_1(y)$ has a
local maximum; and $c = F_1(y*)$.

This problem is equivalent to that shown in Fig.~\ref{fig:Fig3}~c
and \ref{fig:Fig3}~d provided $A^2 >1$, $4B^2<1$. The effective
potential possesses a local minimum and a local maximum at the
left side of the minimum.

Using the results of papers \cite{28:GH92,29:GM96} we can solve
exactly the dynamical problem of the soliton in this potential.

Our analysis shows that if $2B^2(3A^2-1)<1$, then the soliton
feels the barrier in the point $y = 0$. So, if the wave velocity
$w$ is small, the soliton can be carried to the right. On the
other hand, if $2B^2(3A^2-1)>1$ (although the waveform is still as
that shown in Fig.~\ref{fig:Fig3}~c such that $F(y)$ has two
zeroes), the soliton will move to the left, crossing the barrier
even if its center--of--mass is placed in the minimum of the
potential and its initial velocity is zero. The soliton does not
feel the barrier. Even with an infinitesimal wave velocity, the
soliton will not move to the right dragged by the wave. We have
checked this phenomenon numerically.

The effect of a wave shaped as that in Fig.~\ref{fig:Fig4} can be
seen in the following example with exact solution:

%-------------------------------------------------------------------------------------
\begin{equation}
\phi_{tt} + \gamma\phi_{t} - \phi_{xx} -\frac{1}{2}\phi +
\frac{1}{2}\phi^3 =
\frac{A}{\cosh^2\left[\frac{1}{2}\left(\frac{x-wt}{\sqrt{1-w^2}}\right)\right]}.
\label{Eq13:10}
\end{equation}
%-------------------------------------------------------------------------------------

When $A = -\frac{\gamma}{2}\frac{w}{\sqrt{1-w^2}}$, the following
travelling soliton is an exact solution $\phi = \tanh\left[
\frac{1}{2}\left( \frac{x-wt}{\sqrt{1-w^2}} \right)\right]$. Note
that a wave shaped like that in Fig.~\ref{fig:Fig4} can make the
soliton move to the right. However, for a fixed $A$ in
Eq.~(\ref{Eq13:10}), the wave velocity should not exceed a limit.
Otherwise, the soliton can be left behind.

In the case of a sinusoidal wave acting in the framework of the
$\phi^4$ equation:
%-------------------------------------------------------------------------------------
\begin{equation}
\phi_{tt} + \gamma\phi_t - \phi_{xx} - \frac{1}{2}\phi +
\frac{1}{2}\phi^3 =  A \sin\left[ k\left( x - wt\right) \right],
\label{Eq14:11}
\end{equation}
%-------------------------------------------------------------------------------------
we can write the following conditions for the soliton to be
dragged  by the wave to the right: $\frac{\pi}{k}>4$, $w<v_m$,
where $v_m$ is defined by the equation $\frac{\gamma
v_m}{\sqrt{1-v_m^2}}= \frac{\phi_2}{2}$, and $\phi_2$ is the
``middle'' root to the algebraic equation $\phi - \phi^3 = 2A$ as
in Eq.~(\ref{Eq7:DS}).

We have experimented with other waveforms that illustrate the
different behaviors discussed above.

The force $F(x,t)= a \mathrm{tanh} [B(x-wt)]$, with $a<0$ has the
properties of the function  shown in Fig.~\ref{fig:Fig1}b.

%-------------------------------------------------------------------------------------
\begin{figure}
\centering
\includegraphics[width=7.5cm]{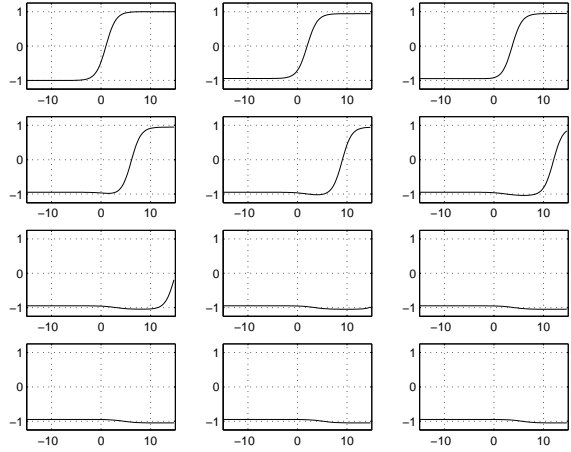}
\caption{A kink--soliton can be pushed by unstable equilibrium.
$F(x,t)=a\tanh[B(x-wt)]$, $\phi(x,0)=2B\tanh[B(x-x_0)]$,
$\phi_t(x,0)=0$, $B = 0.5$, $w = 0.05$, $a = -0.05$, $x_0 = 1.0$,
$\gamma =  0.5$}
\label{fig:FigE1}
\end{figure}
%-------------------------------------------------------------------------------------

Fig.~\ref{fig:FigE1} shows that the soliton can be pushed by this
external perturbation, provided the velocity $w$ is not too large.

\begin{figure}
\centering
\includegraphics[width=7.5cm]{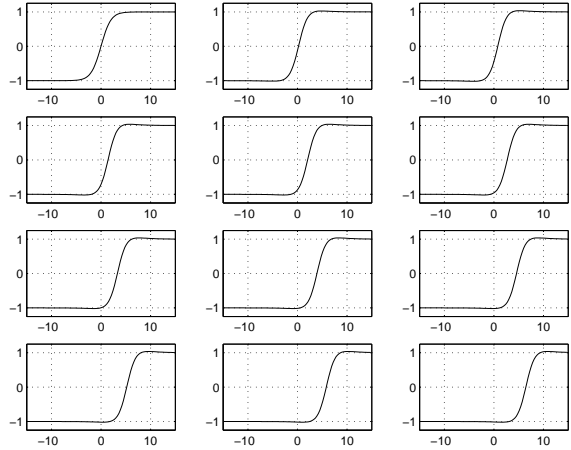}
\caption{The soliton is transported by the perturbation:
$F(x,t)=a\sinh[B(x-wt)]/\cosh^3[B(x-wt)]$,
$\phi(x,0)=2B\tanh(Bx)$, $\phi_t(x,0)=0$, $B = 0.5$, $w = 0.1$,
$a=0.2$, $\gamma = 0.5$}
\label{fig:FigE2}
\end{figure}

In Fig.~\ref{fig:FigE2} we present the behavior of the soliton
under the action of the external perturbation
$F(x,t)=\frac{a\sinh[B(x-wt)]}{\cosh^3[B(x-wt)]}$,
which is of the kind shown in Fig.~\ref{fig:Fig1}c.

%-------------------------------------------------------------------------------------
%-------------------------------------------------------------------------------------

When $w=0.1$, the soliton is transported in a controlled way by
the external wave.

\begin{figure}
\centering
\includegraphics[width=7.5cm]{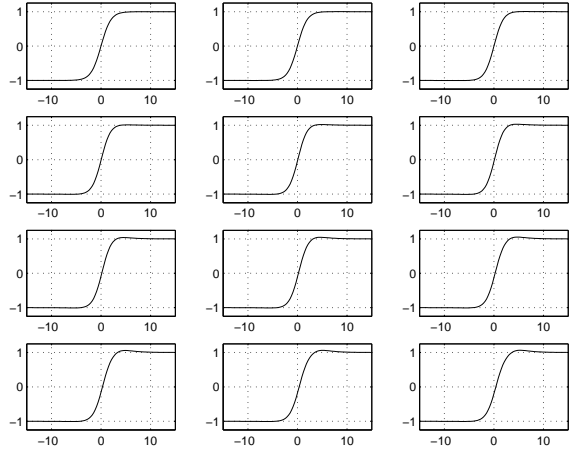}
\caption{The same experiment as in Fig.~\ref{fig:FigE2}. However,
the wave velocity is larger, and the soliton is not transported by
the wave. ($F(x,t)=a\sinh[B(x-wt)]/\cosh^3[B(x-wt)]$,
$\phi(x,0)=2B\tanh(Bx)$, $\phi_t(x,0)=0$, $B = 0.5$, $w = 0.5$,
$a=0.2$, $\gamma = 2.0$)}
\label{fig:FigE3}
\end{figure}

However, already when $w=0.5$, the soliton is abandoned behind.
This can be observed in Fig.~\ref{fig:FigE3}.

%-------------------------------------------------------------------------------------
%-------------------------------------------------------------------------------------

\begin{figure}
\centering
\includegraphics[width=7.5cm]{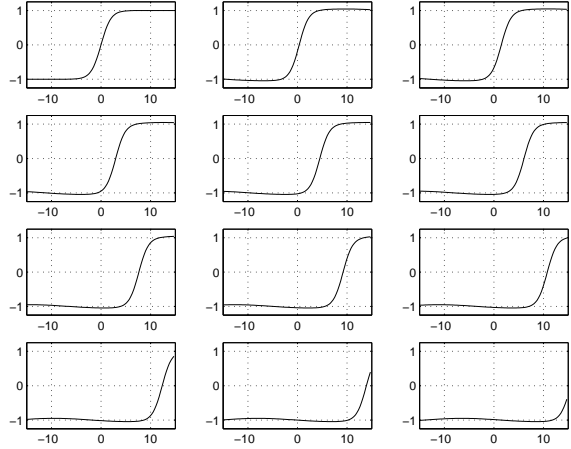}
\caption{A periodic wave can carry the soliton. (
$F(x,t)=A\sin[\kappa x - wt]$, $\phi(x,0)=2B\tanh(Bx)$,
$\phi_t(x,0)=0$, $B = 0.5$, $w = 0.01$, $A=0.05$, $\kappa = 0.2$
$\gamma = 0.5$)}
\label{fig:FigE4}
\end{figure}

A periodic wave like $F(x,t)= A\sin[kx-wt]$ is able to carry the
soliton as is shown in Fig.~\ref{fig:FigE4} where $w = 0.01$. The
values of the other parameter can be found in the figure caption
of Fig.~\ref{fig:FigE4}.

%-------------------------------------------------------------------------------------
%-------------------------------------------------------------------------------------

\begin{figure}
\centering
\includegraphics[width=7.5cm]{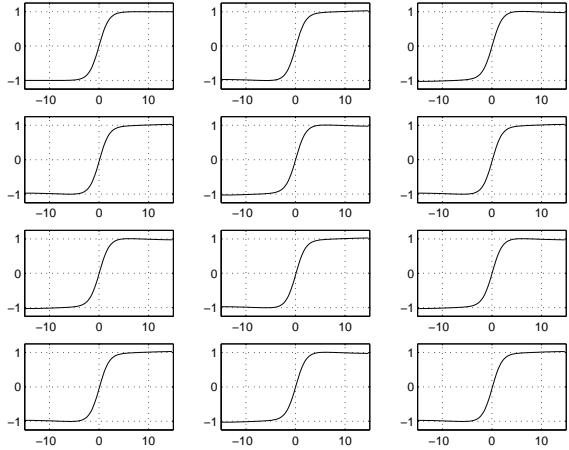}
\caption{The periodic wave must satisfy some combinations in order
to transport the soliton. In this case, for a larger wave
velocity, the soliton is left behind. ($F(x,t)=A\sin[\kappa x -
wt]$, $\phi(x,0)=2B\tanh(Bx)$, $\phi_t(x,0)=0$, $B = 0.5$, $w =
0.9$, $A=0.05$, $\kappa = 0.2$ $\gamma = 0.5$)}
\label{fig:FigE5}
\end{figure}

If the wave velocity is too large (say $w = 0.9$), then the
soliton cannot be carried by the external wave (See
Fig.~\ref{fig:FigE5}).

%-------------------------------------------------------------------------------------
%-------------------------------------------------------------------------------------

%--------------------------------------------------------------------------------
\section{Bubbles in soliton equations}
%--------------------------------------------------------------------------------

Phase transitions
\cite{33:GO99,34:Lan67,35:BL81,36:Au75,37:Col77,38:CC77,39:CBCR,40:GB98,41:FKS92,42:PS79,43:PP75,44:Pa78}
can be described by an equation similar to (\ref{Eq2:KG}):
%-------------------------------------------------------------------------------------
\begin{equation}
\phi_{tt} + \gamma\phi_t - \nabla^2\phi =  - \frac{\partial
U(\phi)}{\partial \phi} + F(x,y,z,t).
\label{Eq15:12}
\end{equation}
%-------------------------------------------------------------------------------------
Recall that  $U(\phi)$ possesses two minima ($\phi_1$ and
$\phi_3$) and a maximum ($\phi_1<\phi_2<\phi_3$).

Suppose for the moment that $U(\phi_1)<U(\phi_3)$. The main
phenomenon for $F\equiv 0$ is that of ``nucleation'' where drops
or bubbles grow or disappear depending on their Gibbs energies
\cite{33:GO99,34:Lan67,35:BL81,36:Au75,37:Col77,38:CC77,39:CBCR,40:GB98,41:FKS92,42:PS79,43:PP75,44:Pa78}.

The relevant issue is the transition from the metastable state
$\phi_3$ to the state $\phi_1$. In phase--transition theory,
researchers  consider the existence of a critical ``germ'' for the
transition. Field configurations $\phi(\mathbf{x},t)$ with a
``radius'' larger than that of the critical germ should grow in
order to produce the transition to the state $\phi_1$.

Let us first discuss this situation in the context of the
one--dimensional equation
%-------------------------------------------------------------------------------------
\begin{equation}
\phi_{tt} + \gamma \phi_t - \phi_{xx} = - \frac{\partial
U(\phi)}{\partial \phi}.
\label{Eq16:13}
\end{equation}
%-------------------------------------------------------------------------------------

\begin{figure}
\centering
\includegraphics[width=7.0cm]{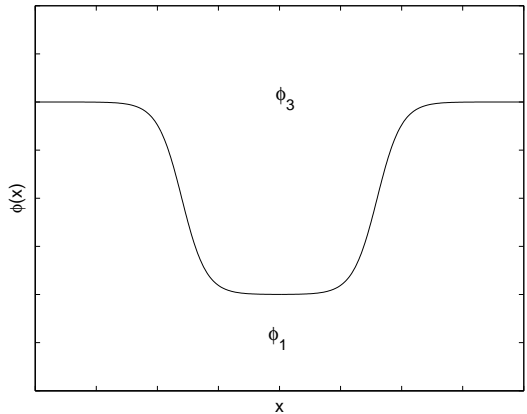}
\caption{A ``bubble'' formed by a kink--antikink pair in
Eq.~(\ref{Eq17:14})}
\label{fig:Fig7}
\end{figure}

For $U(\phi_1)<U(\phi_3)$, it is possible to show (see
Refs.~\cite{33:GO99} and \cite{45:GH87}) that there exists an
unstable stationary critical solution that plays the role of a
critical germ. This solution is represented in
Fig.~\ref{fig:Fig7}.

%-------------------------------------------------------------------------------------
%-------------------------------------------------------------------------------------

This solution can be interpreted as a kink--antikink pair
\cite{45:GH87,46:GES89}.

In general, a kink and an antikink attract each other
\cite{46:GES89,47:MGGLA98}. On the other hand, a constant external
force acts on a kink and an antikink in different directions. So
the solution represented in Fig.~\ref{fig:Fig7} is possible when
the kink and the antikink are at some critical distance where the
attraction force and the external force balance each other
\cite{46:GES89,47:MGGLA98}.

The critical solution is unstable \cite{33:GO99,46:GES89}.
Perturbations of the critical solution can lead to the growth of
the critical ``bubble'' initiating the transition to the state
$\phi = \phi_1$. ``Bubbles'' with a radius smaller than the radius
of the critical solution will disappear, whereas ``bubbles'' with
a radius larger than the radius of the critical solution will grow
until the whole system is in the state $\phi = \phi_1$.

An example can be studied in the equation
%-------------------------------------------------------------------------------------
\begin{equation}
\phi_{tt} + \gamma\phi_t - \phi_{xx} - \frac{1}{2}\phi +
\frac{1}{2}\phi^3 = F_0,
\label{Eq17:14}
\end{equation}
%-------------------------------------------------------------------------------------
where $F_0=\mathrm{const}$.

Note that Eq.~(\ref{Eq17:14}) can be written in the form
(\ref{Eq16:13}), where
%-------------------------------------------------------------------------------------
\begin{equation}
U(\phi) = \frac{1}{8}\left(\phi^2-1\right)^2 - F_0\phi.
\label{Eq18:15}
\end{equation}
%-------------------------------------------------------------------------------------

If $F_0<0$, then $U(\phi_1)< U(\phi_3)$. Eq.~(\ref{Eq17:14})
possesses a bell--shaped stationary solution as that shown in
Fig.~\ref{fig:Fig7}. This solution is unstable
\cite{45:GH87,46:GES89}. Similar but different bell--shaped
initial states will evolve either to the state $\phi=\phi_1$ or
the state $\phi = \phi_3$. If the kink and the antikink that form
the one--dimensional ``bubble'' are too close, then the bubble
will collapse. If the distance between the kink and the antikink
is larger than the ``width'' of the critical germ, then the
``bubble'' will grow.

If $F_0 = 0$, all the initial bubbles will collapse.

Can we stabilize the bubbles of phase $\phi = \phi_1$ inside the
``sea'' of pase $\phi=\phi_3$ using external perturbations?

\begin{figure}
\centering
\includegraphics[width=7.0cm]{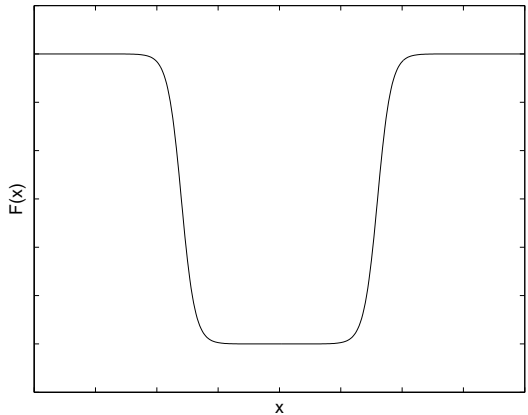}
\caption{Inhomogeneous external perturbation able to stabilize the
bubble in Eq.~(\ref{Eq19:16})}
\label{fig:Fig8}
\end{figure}

Let us analyze what happens when a kink--antikink pair as that
shown in Fig.~\ref{fig:Fig7} is in the presence of an external
perturbation like that shown in Fig.~\ref{fig:Fig8}

%-------------------------------------------------------------------------------------
%-------------------------------------------------------------------------------------

Note that these graphs look similar but they represent very
different physical quantities.

\begin{figure}
\centering
\includegraphics[width=7.0cm]{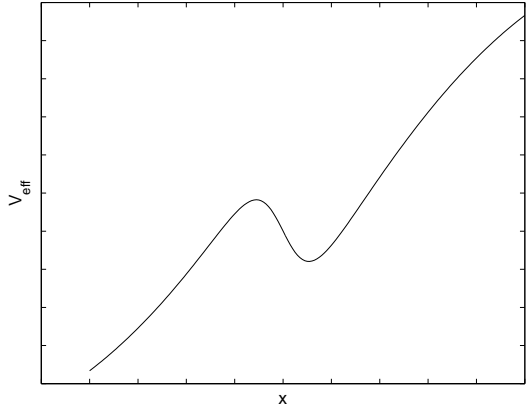}
\caption{Effective potential for the kink that forms the
``bubble'' shown in Fig.~(\ref{fig:Fig7}) in the presence of
perturbation shown in Fig.~(\ref{fig:Fig8})}
\label{fig:Fig9}
\end{figure}

\begin{figure}
\centering
\includegraphics[width=7.0cm]{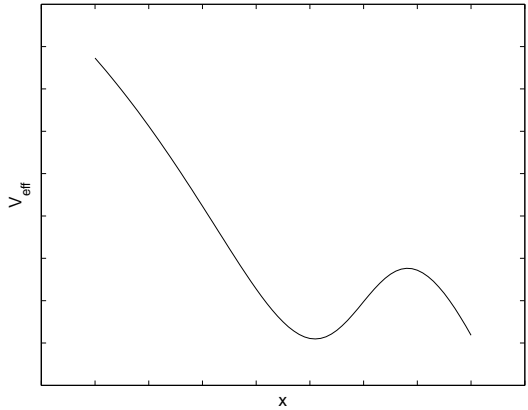}
\caption{Effective potential for the antikink that forms the
``bubbles'' shown in Fig.~(\ref{fig:Fig7}) in the presence of
perturbation shown in Fig.~(\ref{fig:Fig8})}
\label{fig:Fig10}
\end{figure}

If the kink of Fig.~\ref{fig:Fig7} is close to the ``right'' zero
of function $F(x)$, it will ``feel'' an effective potential well
(see Fig.~\ref{fig:Fig9}). The same will happen to the antikink if
it is close to the ``left'' zero of $F(x)$ in Fig.~\ref{fig:Fig8}
(See the potential in Fig.~(\ref{fig:Fig10}).)

%-------------------------------------------------------------------------------------
%-------------------------------------------------------------------------------------

%-------------------------------------------------------------------------------------
%-------------------------------------------------------------------------------------

If the potential barrier is sufficiently high, the attraction
force between the kink and the antikink will not lead to the
``bubble'' collapse. The attraction force decays rapidly
(exponentially) with the distance, so for large ``bubbles''
practically ``any'' $F(x)$ with the shape of Fig.~\ref{fig:Fig8}
is good for stabilizing the ``bubble''.

On the other hand, even small bubbles can be stabilized using an
external perturbation $F(x)$ with two zeroes $x_1$ and $x_2$ (as
in Fig.~\ref{fig:Fig8}) such that $\left| \left(\frac{\mathrm{d}
F(x)}{\mathrm{d}x}\right)_{x=x_{1,2}} \right|$ is sufficiently
large. That is, the change of sign in $F(x)$ should be
sufficiently sharp.

We have checked numerically the existence of the stable
``bubbles''.

Let us experiment with the following equation
%-------------------------------------------------------------------------------------
\begin{equation}
\phi_{tt} + \gamma\phi_t - \phi_{xx} - \frac{1}{2}\phi +
\frac{1}{2}\phi^3 = F(x),
\label{Eq19:16}
\end{equation}
%-------------------------------------------------------------------------------------
where
%-------------------------------------------------------------------------------------
\begin{equation}
F(x) = A \left\{ \tanh\left[ B(x-d) \right] - \tanh\left[ B(x+d)
\right] + \epsilon \right\}
\label{Eq20:17}
\end{equation}
%-------------------------------------------------------------------------------------

For instance, we have established  that the perturbations
(\ref{Eq20:17}), with the parameters $A=1$, $B = 0.2$, $d= 5$,
$\epsilon = 0.7$, can sustain a stable bubble.

An initial configuration for $\phi(x,t)$ where the kink and the
antikink are not exactly in the zeroes of $F(x)$ will, anyway,
tend to the asymptotically stable bubble. That is, there is a
bubble which is a spatiotemporal attractor. All close initial
configurations will tend (for $t\rightarrow\infty$), to the
asymptotically stable bubble.

This phenomenon occurs in the case that a bubble is considered as
an initial configuration. That is, a bubble had been created
before the ``force'' $F(x)$ is applied. However, it is interesting
to remark that for sufficiently large $A$ and $B$, the
perturbation itself can create the bubble. That is, there is no
need for a preexisting bubble. The bubble is created by the
perturbations and, then, the perturbation can sustain the
stabilized bubble.

If $A$ is too small, it is possible that  $F(x)$ cannot sustain a
bubble. An example is produced with the following parameters
$A=0.25$, $B=1.0$, $d=1.0$, $\epsilon = 0.8$. All initial
conditions lead to the bubble collapse.

Once we have an external perturbation $F(x)$ able to sustain a
stable bubble, we can transport the bubble if we can move $F(x)$
as a whole. This phenomenon is in the same spirit of all the
theory discussed above concerning the soliton transport.

In the case of the bubble, as in many of the discussed situations
with solitons, there is  a limit for the velocity of the
perturbation as a whole. When the shape represented in
Fig.~\ref{fig:Fig8} is moved too fast, the bubble can be left
behind, and then it will collapse.

%--------------------------------------------------------------------------------
\section{Three dimensional bubbles}
%--------------------------------------------------------------------------------

What about the bubbles in three dimensions?

Let us concentrate on solutions with radial symmetry.

The stationary solutions of Eq.~(\ref{Eq15:12}) ($F(x,y,z,t)\equiv
0$), satisfy the equation
%-------------------------------------------------------------------------------------
\begin{equation}
\phi_{rr} + \frac{2}{r}\phi_r - \frac{\partial U(\phi)}{\partial
\phi}=0.
\label{Eq21:18}
\end{equation}
%-------------------------------------------------------------------------------------

It is possible to prove \cite{33:GO99} that  (for
$U(\phi_1)<U(\phi_3)$) there exists a solution with a minimum in
$r=0$ and that asymptotically it tends to $\phi_3$ as
$r\rightarrow\infty$. This solution corresponds to the critical
germ discussed above. Again as before, this critical solution is
unstable.

In general, exact solutions for the three--dimensional case of the
nonlinear equation (\ref{Eq21:18}) are very difficult to find
\cite{33:GO99}. Nevertheless, we have developed an
inverse--problem approach that allows us to find exact solutions
to a class of topologically equivalent systems.

For instance, a wealth of work \cite{33:GO99} has been dedicated
to find a solution to equations of type
%-------------------------------------------------------------------------------------
\begin{equation}
\phi_{rr} + \frac{2}{r}\phi_r - B^2\phi + D^2\phi^3 + T(\phi)=0,
\label{Eq22:19}
\end{equation}
%-------------------------------------------------------------------------------------
where $T(\phi)$ possesses only terms of higher order than $3$.
This equation is important in field theory.

We have found that function
%-------------------------------------------------------------------------------------
\begin{equation}
\phi = \frac{A\tanh (Br)}{r\cosh (Br)}
\label{Eq23:20}
\end{equation}
%-------------------------------------------------------------------------------------
is the exact solution to equation
%-------------------------------------------------------------------------------------
%\begin{equation*}
$\phi_{rr} + \frac{2}{r} \phi_r - B^2\phi + \frac{1}{A^2}\phi^3 +
T(\phi)=0,$
%
%\end{equation*}
%-------------------------------------------------------------------------------------
where $T(\phi)$ contains only terms of order higher than $3$.

Likewise, it can be proved that function
%-------------------------------------------------------------------------------------
\begin{equation}
\phi = \phi_3 - \frac{A\tanh (Br)}{r \cosh(Br + D)}
\label{Eq24:21}
\end{equation}
%-------------------------------------------------------------------------------------
is the critical germ solution of an equation of type
(\ref{Eq21:18}), where $U(\phi)$ possesses all the properties that
we need: two minima and a maximum, $U(\phi_1)<U(\phi_3)$, etc.

As in the one--dimensional case, small bubbles will collapse, and
bubbles with a radius larger than that of the critical germ will
grow.

In the context of the three--dimensional equation:
%-------------------------------------------------------------------------------------
\begin{equation}
\phi_{tt} + \gamma\phi_t - \nabla^2\phi - \frac{1}{2}\phi +
\frac{1}{2}\phi^3 = 0,
\label{Eq25:22}
\end{equation}
%-------------------------------------------------------------------------------------
all the bubbles are non--stationary. However, we should say that
very large bubbles are quasi--stationary because the interaction
between the walls is so small that (although they should collapse
eventually) their life--time is very large.

What about inhomogeneous perturbations as in Eq.(\ref{Eq15:12})?

Suppose we have an initial configuration $\phi(\mathbf{x},t=0)$
with the shape of a spherical bubble of phase $\phi = \phi_1$
inside a ``sea'' of phase $\phi_3$. This means that there is a
spherical spatial sector where $\phi\approx \phi_1$, and, outside
this sector, $\phi\approx\phi_3$. Space zones where $F(x,y,z)<0$
will ``try'' to expand the bubble. Space zones where $F(x,y,z)\geq
0$ will ``try'' to make the bubble collapse.

In fact, as in the one--dimensional case, we can stabilize a
spherical bubble using an inhomogeneous external perturbation
$F(x,y,z)$ such that $F(x,y,z)<0$ for points $(x,y,z)$ that are
inside the bubble and $F(x,y,z)>0$  for points $(x,y,z)$ outside
the bubble.

To ensure the stability of the bubble, the sign change of
$F(x,y,z)$ should be sharp. We have an exact solution.

Let us define the equation
%-------------------------------------------------------------------------------------
\begin{equation}
\phi_{tt} + \gamma\phi_t - \nabla^2\phi -\frac{1}{2}\phi +
\frac{1}{2}\phi^3=F(r),
\label{Eq26:23}
\end{equation}
%-------------------------------------------------------------------------------------
where
%-------------------------------------------------------------------------------------
\begin{eqnarray}
F(r) &=& \frac{1}{2}A\left( A^2-1 \right)\left\{ \tanh\left[
\frac{A}{2}\left(r-R\right)\right] \right.\nonumber\\
&&\left.- \tanh\left[ \frac{A}{2}(r+R) \right] \right\}\nonumber\\
&& + \frac{3}{2}\left\{ \left( \phi_1 + \phi_2 \right)\left[
\phi_1\phi_2 + A\left( \phi_1 + \phi_2 \right) + A^2 \right]
\right\}\nonumber\\
&& -\frac{2}{r}\left[ \phi_{1r} + \phi_{2r} \right] - \frac{1}{2}A
+ \frac{1}{2}A^3,\nonumber
\end{eqnarray}
%-------------------------------------------------------------------------------------
$r=\sqrt{x^2+y^2+z^2}$, $\phi_1 = A\tanh\left[ B(r-R) \right]$,
$\phi_2 = -A\tanh\left[ B(r+R) \right] $; $A$, $B$ and $R$ are
real parameters.

Here $F(r)$ has the desired property that $F(r)<0$ in a spherical
sector around $r=0$, and $F(r)>0$ outside this spherical sector.

There is an exact bubble solution to Eq.~(\ref{Eq26:23}):
%-------------------------------------------------------------------------------------
\begin{equation}
\phi = A \left\{ \tanh \left[ \frac{A}{2}\left(r-R\right) \right]
- \tanh\left[\frac{A}{2}\left( r+R \right)\right] + 1 \right\}.
\label{Eq27:24}
\end{equation}
%-------------------------------------------------------------------------------------

But we can say more. For $R>5$, $A>1$, this bubble solution is
stable. Manipulating $F(r)$ we can control this bubble. In fact,
we can move this bubble to another point in space.

%--------------------------------------------------------------------------------
\section{Bubbles in phase transitions}
%--------------------------------------------------------------------------------

Many works have been dedicated to the nucleation process in phase
transitions
\cite{33:GO99,34:Lan67,35:BL81,36:Au75,37:Col77,38:CC77,39:CBCR,40:GB98,41:FKS92,42:PS79,43:PP75,44:Pa78,45:GH87}.
Here we wish to focus on the works based on the theory of
relaxation of metastable states \cite{42:PS79,43:PP75,44:Pa78}.

In Refs.\cite{42:PS79,43:PP75,44:Pa78}, the nucleation in the
metastable phase near the critical point of a thermodynamic system
is described as a relaxation of a metastable state of the
order--parameter field.

A metastable state of matter can be produced in a first--order
phase transition. Its relaxation into a thermodynamically stable
state is facilitated by the formation of a critical nucleus of a
new phase.

The system is described by a scalar field $\phi(\mathbf{x},t)$ for
which a relaxation equation is obtained:
%-------------------------------------------------------------------------------------
\begin{equation}
\frac{1}{\Gamma_n}\frac{\partial \phi}{\partial t}= -\frac{\delta
H}{\delta\phi} + f_{ext},
\label{Eq28:25}
\end{equation}
%-------------------------------------------------------------------------------------
where $H[\phi]$ is the system free--energy which is taken in the
Landau form:
%-------------------------------------------------------------------------------------
\begin{equation}
H[\phi] = \frac{1}{2}\int \left\{ c\left( \nabla\phi \right)^2 +
\mu\phi^2 + \frac{1}{2}g\phi^4-2h\phi
\right\}\mathrm{d}\mathbf{x}.
\label{Eq29:26}
\end{equation}
%-------------------------------------------------------------------------------------

Here $\Gamma_n$ is a kinetic coefficient and $f_{ext}$ is an
external perturbation.

At $h=0$ and $\mu = \mu_c<0$, there is a critical point in the
system. The line $h=0$ ($\mu<\mu_c$) is a line of first--order
phase transition.

The Eq.~(\ref{Eq28:25}) is an overdamped version of the $\phi^4$
equation discussed above. This equation can be written in
dimensionless units as our equations (\ref{Eq2:KG}). The solution
we have been investigating describes a nucleus of phase $\langle
\phi \rangle = -1$ in phase $\langle \phi \rangle = 1$. At $h<0$,
the nuclei with small radius attenuate, while the nuclei with
large radius increase. The phase $\langle \phi \rangle = 1$ at
$h<0$ is metastable--unstable to formation of a nucleus with a
large radius.

The parameters of the system (\ref{Eq28:25}--\ref{Eq29:26})  for
the liquid--vapor first--order phase transition depend on the
thermodynamics quantities, e.g., pressure and temperature
\cite{43:PP75,44:Pa78}.

If an external perturbation (say a laser beam) creates the
thermodynamics conditions for a first--order transition (e.g.,
$h<h_c<0$ in the context of system (\ref{Eq28:25}--\ref{Eq29:26}))
in a small zone of a liquid mass, then a vapor bubble could be
generated and sustained.

We should say here that in the framework of equation
%-------------------------------------------------------------------------------------
\begin{equation}
\gamma\phi_t -\nabla^2 \phi - \frac{1}{2}\phi + \frac{1}{2}\phi^3
=  F(x,y,z),
\label{Eq30:27}
\end{equation}
%-------------------------------------------------------------------------------------
in order to create a bubble able to increase to a macroscopic
radius, $F(x,y,z)$ must take negative values in a limited spatial
zone in such a way that $F(x,y,z)<-\frac{1}{3\sqrt{3}}$. On the
other hand, once the bubble has been generated, the perturbation
$F(x,y,z)$ necessary to sustain the bubble should be ``negative''
inside the bubble and positive ``outside'', but the absolute value
of $F(x,y,z)$ in the ``negative'' part should not be so high. In
other words, the ``intensity'' of the perturbation needed for
generating the bubble is larger than the needed ``intensity'' for
stabilizing the bubble.

%-------------------------------------------------------------------------------------
%EXPERIMENTS
%-------------------------------------------------------------------------------------

%--------------------------------------------------------------------------------
\section{Real experiments: Bubbles induced and controlled by lasers}
%--------------------------------------------------------------------------------

Now we will discuss several experiments about bubble formation in
a weak absorbing liquid and its subsequent trapping under the
action of a cw laser beam of relatively low power.

The generation of bubbles by intense light pulses was studied in
the papers \cite{48:La74,49:GD85}. On the other hand, the
pioneering works about laser bubble trapping can be found in
Refs.~\cite{50:Ma92,51:MA93}. Here we present a review of previous
\cite{48:La74,49:GD85} and very recent experiments concerning new
effects related to laser bubble trapping, in the light of the
theory we are developing.

High--intensity light induces a local first--order liquid--vapor
phase transition. If light power is reduced after the bubble
formation, the bubble remains trapped. If the intensity is very
low then the bubble collapses.

\begin{figure}
%\centering
\includegraphics[width=7.0cm]{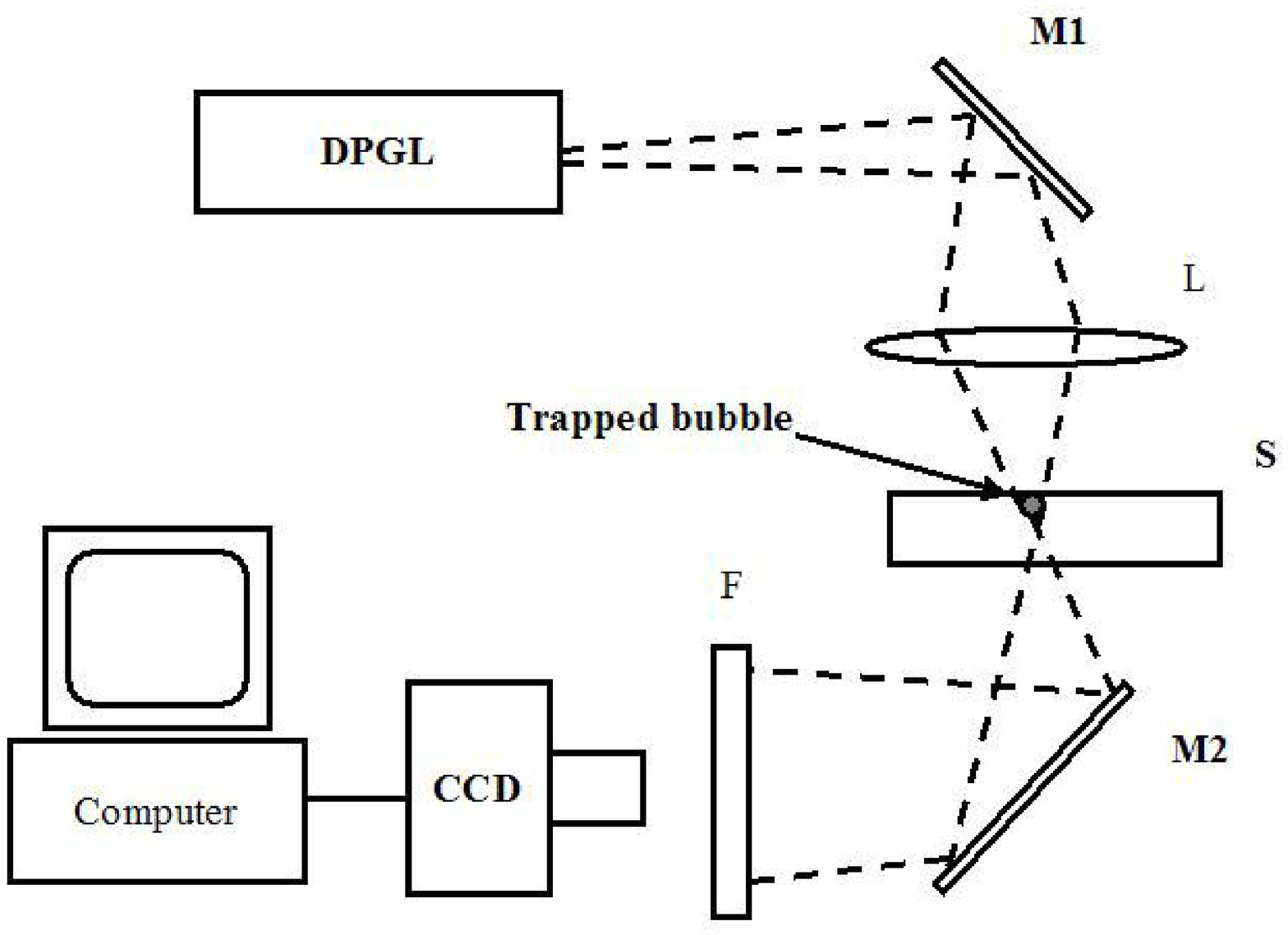}
\caption{Experimental Setup}
\label{fig:FigSetup}
\end{figure}

The experimental setup is shown in Fig.~\ref{fig:FigSetup}.

%-------------------------------------------------------------------------------------
%-------------------------------------------------------------------------------------

In Figure \ref{fig:FigSetup} we show the experimental set-up used
for the laser bubble generation and subsequent trapping. A CW 200
mW diode pumped green laser (l=532 nm) is directed vertically down
toward a horizontally located sample cell using the mirror M1. A
15-cm focal length lens L focuses this light onto the sample. By
moving this lens we can change the radius of the focal spot. The
sample is a 4-mm thick glass cell containing an iodine ethanol
solution of high concentration (1 to 10 ml). The focused beam
heats this solution, generating locally a small bubble at power
levels larger than 60 mW. Following the light absorption a
distribution of temperatures is induced in the focal region and
the bubble surface. Because the surface tension depends upon the
temperature, a spatial distribution of surface tension is
generated on the bubble. The generated gradient of surface tension
induces a force that traps the bubble at the point of maximal
temperature (focal point). This generated force together with the
floatation Archimedes force make the bubble sticks to the upper
surface of the cell. By moving the laser beam using the mirror M1,
the bubble follows the beam spot demonstrating the actual bubble
trapping by the laser beam. The bubble can be observed using the
mirror M2 and a CCD camera connected to a computer.

The used liquids were ethanol, benzene, and ethylene glycol. The
liquids were colored with iodine or a dye in order to increase
absorption.

We will focus here on the bubble behavior. There are several
interference phenomena that will not be discussed here.

\begin{figure}
\centering
\includegraphics[width=4.0cm]{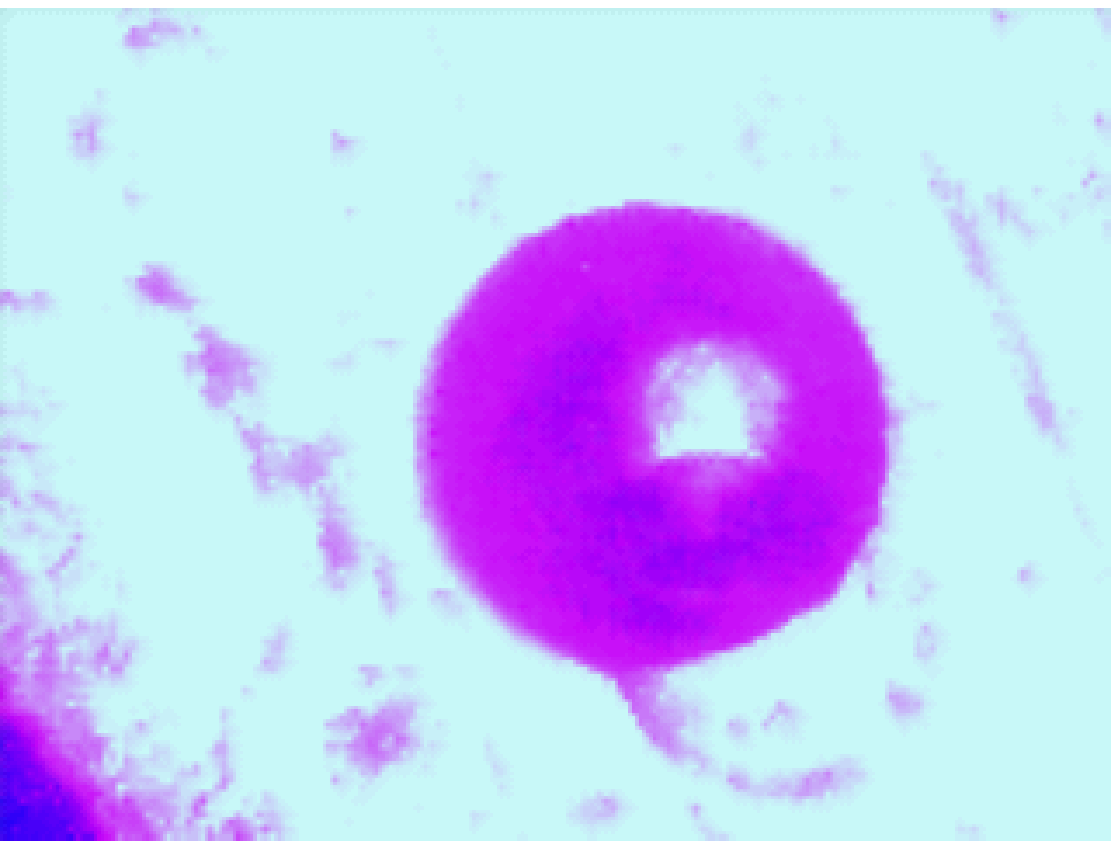}
\includegraphics[width=4.0cm]{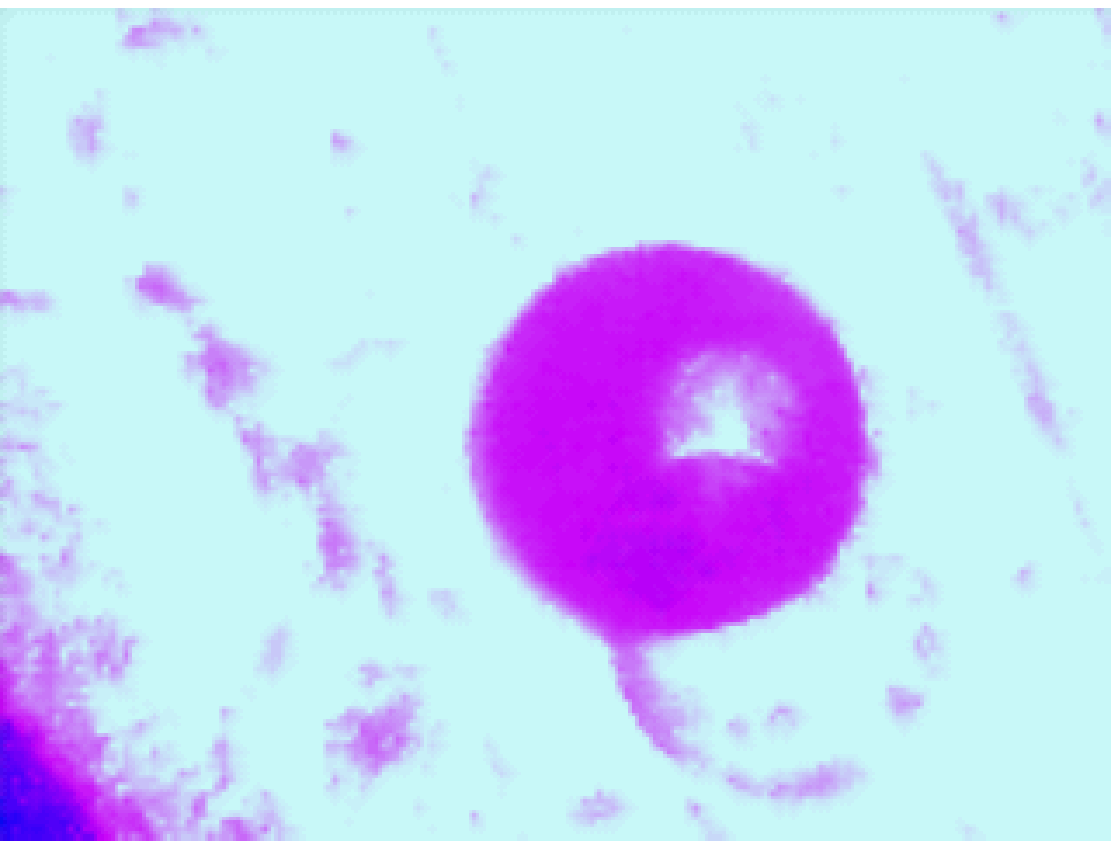}
\includegraphics[width=4.0cm]{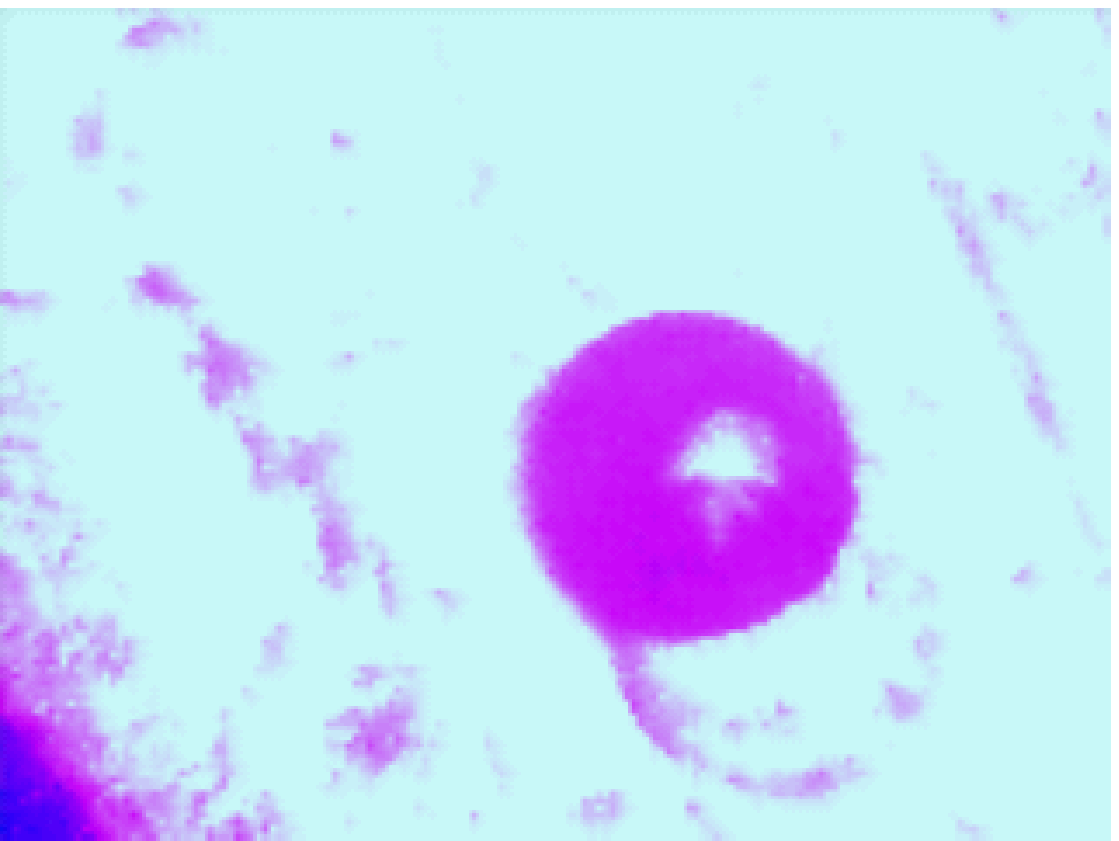}
\caption{Trapped bubble using a laser. The laser power is
$1.7$~mW (left), $0.5$~mW (center) and $0.15$~mW (right). The
bubble's diameters depend on the laser power.}
\label{fig:FigBub1}
\end{figure}

\begin{figure}
\centering
\includegraphics[width=4.0cm]{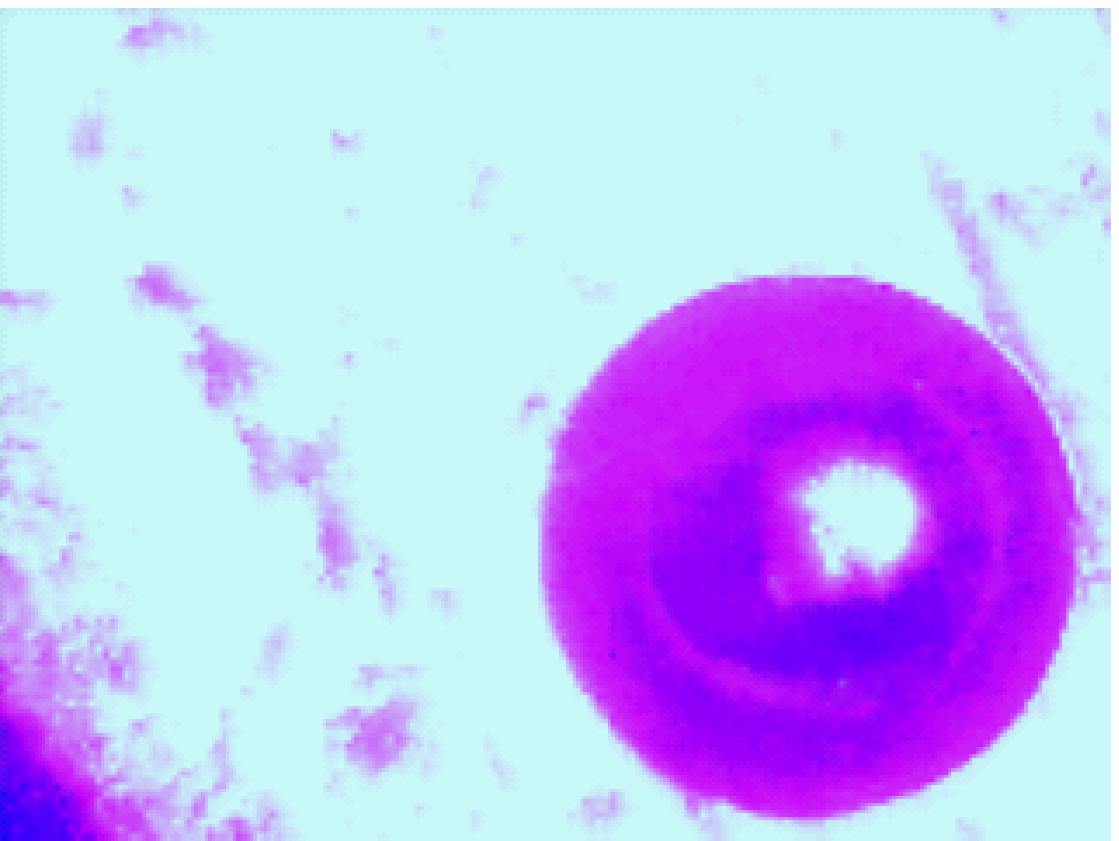}
\includegraphics[width=4.0cm]{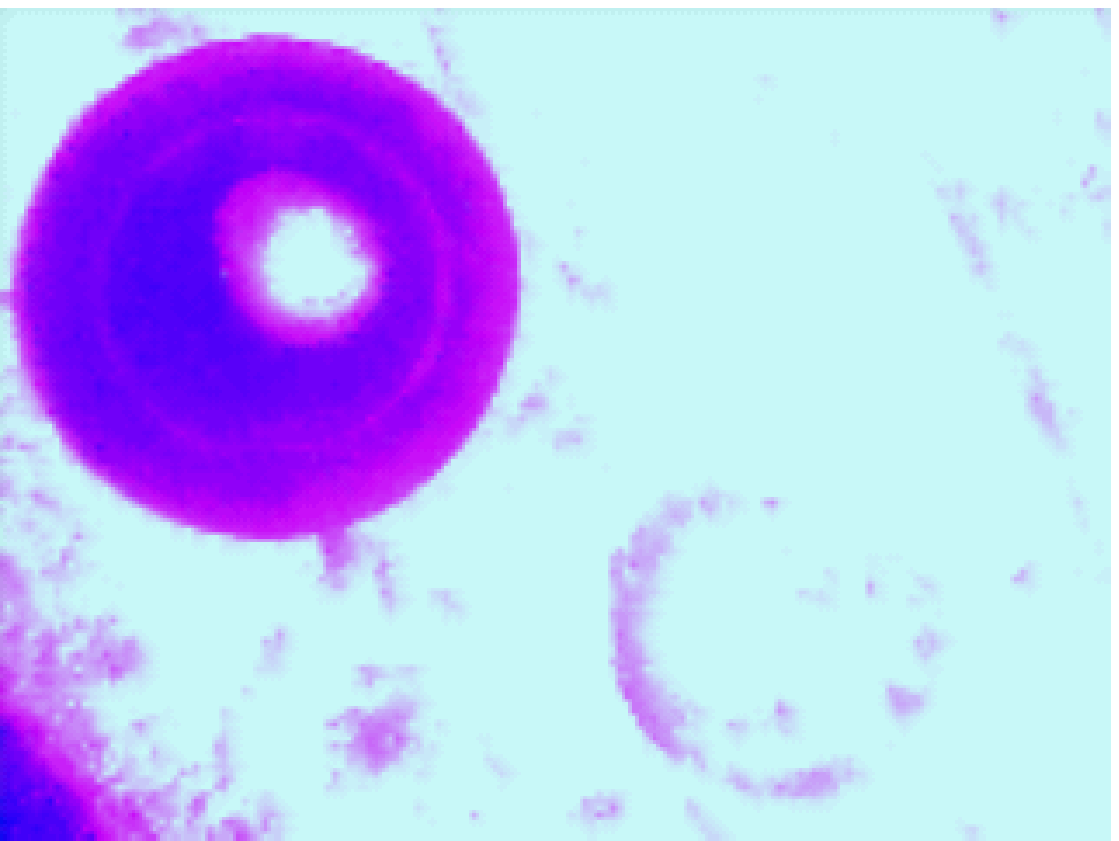}
\caption{Transport of the bubble: (left) The bubble is shown at
its initial position. (right) The bubble is moved to the left
superior corner}
\label{fig:FigBub2}
\end{figure}

\begin{figure}
\centering
\includegraphics[width=4.0cm]{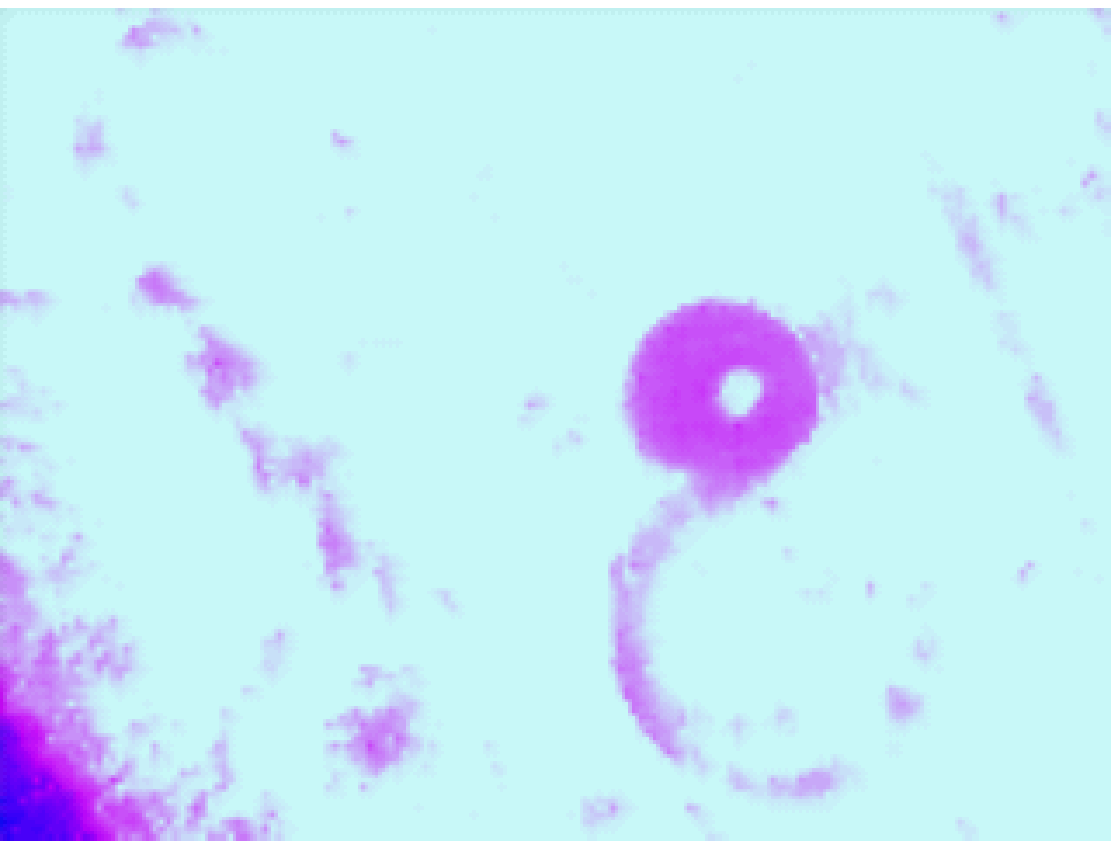}
\includegraphics[width=4.0cm]{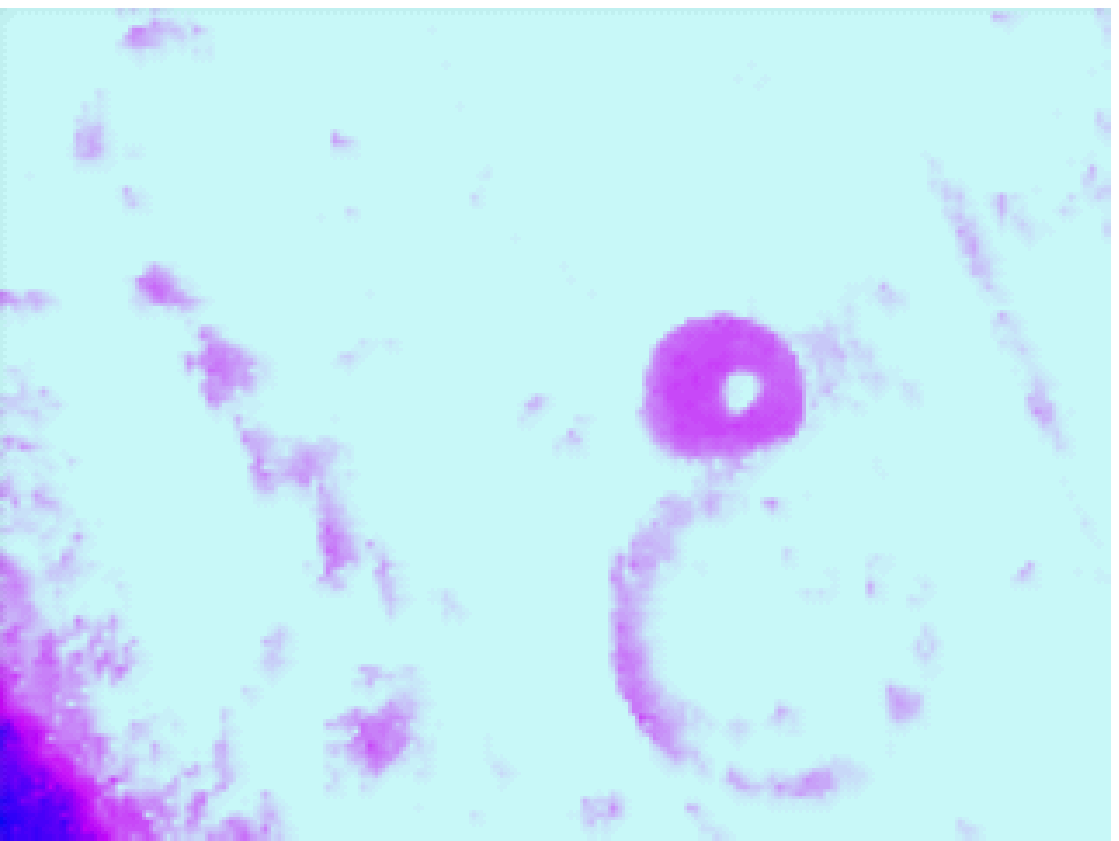}
\includegraphics[width=4.0cm]{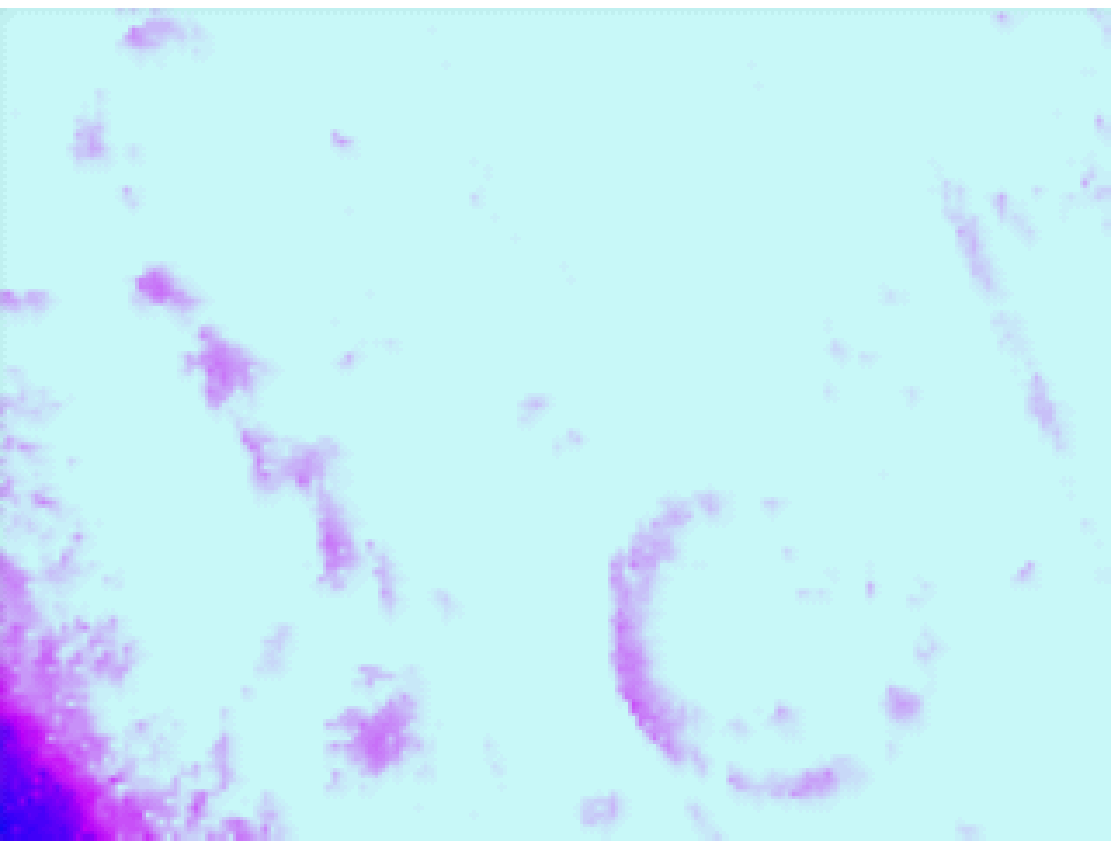}
\caption{The bubble can collapse when the laser power is small.}
\label{fig:FigBub3}
\end{figure}

In Fig.~\ref{fig:FigBub1}--\ref{fig:FigBub3} pictures of the
trapped bubbles are shown.

%-------------------------------------------------------------------------------------
%-------------------------------------------------------------------------------------

In one of the experiments the bubble is generated at a laser power
of $70$~mW. The bubble size decreases with decreasing power.

These phenomena can be described in terms of thermal gradients and
forces \cite{51:MA93}. However, we wish to stress the universal
character of the phenomena related to bubble formation and
transport. These effects can occur in many other physical systems
with phase transitions, solitons, domain walls and bubbles of one
phase inside other phase,

The bubble can be controlled by the laser.

If after creating the bubble, the light beam is defocused in such
a way that its spot size becomes much larger than the radius of
the generated bubble, then changes of the bubble radius occur
until this radius is again stabilized. This can be explained using
the interpretation discussed above where the bubble is described
as trapped inside a stable equilibrium position or potential well.

The bubbles are created with a laser power of $70$~mW. Then the
power is decreased about $10$ times.

In Fig.~\ref{fig:FigBub1} we can observe a trapped bubble with a
diameter $0.4$~mm using a laser power $1.7$~mW. When the laser
power is $0.15$~mW the bubble diameter is $0.2$~mm.

Fig.~\ref{fig:FigBub2} shows the transport of the bubble using the
laser beam. With the help of a mirror the light beam is moved. The
trapped bubble can be moved as the hight beam is moved slowly. If
the motion of the light beam is too fast, then the bubble is left
behind and then it collapses.

In Fig.~\ref{fig:FigBub2} (left),  the bubble is shown at is
initial position. Then it is moved to the left superior corner
using the laser beam.

%-------------------------------------------------------------------------------------
%-------------------------------------------------------------------------------------

Fig.~\ref{fig:FigBub3} shows the collapse of the bubble while the
light power is maintained in $0.1$~mW.

%-------------------------------------------------------------------------------------
%-------------------------------------------------------------------------------------

%-------------------------------------------------------------------------------------
\section{Conclusions}
%-------------------------------------------------------------------------------------

We have investigated generalized nonlinear Klein--Gordon equations
perturbed by moving inhomogeneous external perturbations.

We have shown that a kink--soliton can be transported by an
external wave, provided the shape of the wave and its velocity
satisfy certain conditions. For instance, a moving external
perturbation $F(x-vt)$, where function $F(y)$ possesses a zero
$y_0$ such that $\left.\frac{\partial F}{\partial
y}\right|_{y=y_0}>0$, can carry a kink--soliton provided its
velocity does not exceed certain limit. Figures \ref{fig:Fig1},
\ref{fig:Fig3} and \ref{fig:Fig5} show examples of these
transporting inhomogeneous perturbations. Sinusoidal waves as that
shown in Fig.~\ref{fig:Fig5} can also be successful in
transporting a kink--soliton under certain conditions.

We have also determined the necessary conditions to be satisfied
by an external perturbation to stabilize a bubble of one phase
inside another.

If this stabilizing  inhomogeneous external perturbation is moved,
then the bubble can be transported to another space point.

We have presented the results of real experiments with
laser--induced vapor bubbles in liquids. The bubbles can be
created, trapped, stabilized and transported by a laser beam of
relatively low power.

The investigated equations are very general. Besides, the concepts
and results that lead to the existence of the reported phenomena
are also universal. So we believe that these phenomena can be
observed in many other physical systems. For instance, the
stabilization and transport of bubbles using an external
perturbation can occur in other systems undergoing a
phase--transition. So the mentioned bubbles (or drops) can be
superconducting, ferromagnetic, superfluid, etc.

On the other hand, the discussed transport of solitons and
kink--antikink bubbles can be very important in spin systems,
molecular motors, charge--density waves, ferroelectric materials
and Josephson junctions. In particular, in long Josephson
junctions the kink--solitons are fluxons. These fluxons can be
controlled using external currents \cite{26:KM89,31:GBG99} which
can be very relevant in new communication and computer
technologies where the fluxons are used as very stable information
units.

Moreover, the kinks are examples of topological defects. Many
results about the kinks can be generalized to other topological
defects as vortices and spiral waves which are being studied
intensively in countless systems nowadays
\cite{52:CH93,53:AK02,54:BO99,55:ZQ00}.

%--------------------------------------------------------------------------------
%--------------------------------------------------------------------------------
%--------------------------------------------------------------------------------

%-------------------------------------------------------------------------------------

%-------------------------------------------------------------------------------------
\end{document}